\ifx \doiurl    \undefined \def \doiurl#1{\href{http://dx.doi.org/#1}{\textsf{DOI}}}\fi
\ifx \adsurl    \undefined \def \adsurl#1{\href{http://adsabs.harvard.edu/abs/#1}{\textsf{ADS}}}\fi
\ifx \arxivurl  \undefined \def \arxivurl#1{\href{http://arxiv.org/abs/#1}{\textsf{arXiv}}}\fi

\documentclass[pdftex,12pt]{article}
\usepackage{natbib}
\usepackage[T1]{fontenc}
\usepackage[english]{babel}
\usepackage[breaklinks,colorlinks]{hyperref}
\hypersetup{
citecolor=blue,
urlcolor=blue
}
\usepackage{graphicx}        
\usepackage{color}           
\def\arcsec{$^{\prime\prime}$}
\usepackage{amssymb}

\usepackage{textcomp}
\usepackage{gensymb}
\usepackage[font={small}]{caption}

\title{\vskip-2cm
Influence of the Magnetic Field on Oscillation Spectra in Solar Faculae}

\author{A.A.~Chelpanov, N.I.~Kobanov, and D.Y.~Kolobov\\
    \small{Institute of Solar-Terrestrial Physics} \\
    \small {of Siberian Branch of Russian Academy of Sciences, Irkutsk, Russia} \\
    \small {email: \url{chelpanov@iszf.irk.ru}}
    }
\date{\small{[
{This article was firstly published in {\textit{Solar Physics}} \href{https://dx.doi.org/10.1007/s11207-014-0623-6}{DOI}}]}}

\begin{document}
\maketitle
\begin{abstract}
In this work, we studied oscillation parameters in faculae above magnetic knots and in the adjacent to them areas. Using SDO data we analysed oscillations in magnetic strength, Doppler velocity, and intensity signals for the lower photosphere, and in intensity for the higher levels. We found that in the magnetic field strength oscillation spectra in magnetic knots, peaks at a frequency of about 4.8\,mHz appear, while there are no such frequencies in the adjacent facular patches of a moderate field strength. On the contrary, Doppler velocity photospheric oscillation spectra are similar for these types of regions: in both cases, the significant peaks are in the 2.5--4.5\,mHz range, though the oscillations in magnetic knots are 2--3 times weaker than those at the facular periphery. At the upper photosphere, the dominant frequencies in magnetic knots are 0.5--1\,mHz higher than in the medium-field regions. The transition region oscillations above magnetic knots mainly concentrate in the 3--6\,mHz range, and those above moderate-field patches concentrate below 3\,mHz.

\end{abstract}

\section{Introduction}
     \label{S-Introduction}
\par Studying waves in the solar atmosphere can help answer the question of energy transfer to the outer layers of the Sun. Also, the observed characteristics of waves can provide information on the physical conditions at different heights of the atmosphere \citep{moortel, stepanov}. Oscillations in faculae have been of special interest for 50 years \citep{orrall, Howard}, since faculae are the most abundant solar magnetic activity manifestation, and they occupy a considerable fraction of the solar surface.

\par The characteristics of the magnetic field in active regions have been linked to the observed oscillation characteristics. For example, in sunspots, high-frequency oscillations tend to concentrate in the central part of a spot, where magnetic field lines are closest to vertical, and low-frequency oscillations are observed in the outer sunspot regions, where the magnetic field is inclined \citep{Jefferies, McIntosh, Kobanov11a, reznikova, kobanov13}. Oscillations in microwave data are observed above regions with strong magnetic field \citep{abramov}.

\par Faculae as well are characterized by different strength and inclination of the magnetic field, though the distribution of these parameters are more chaotic than those in sunspots (\citealp{ishikawa}; \citealp{pillet}; \citealp{guo}). There are, however, patterns in the oscillation observation results: low-frequency oscillations increase at the facula boundaries, while three- and five-minute oscillations are observed within faculae \citep{kobanov11}. \citet{wijn} registered waves with three-minute period propagate in the central facular chromosphere, and five-minute waves propagate at the peripheral regions. The influence of the magnetic field on oscillation characteristics in faculae was studied by \citet{khomenko, kostik12, kostik13}. \citet{kostik13} noted increase in the observed period in the regions with high magnetic field strength. \citet{turova} reported a strong five-minute period in the chromosphere of a facula at the base of a coronal hole, and decrease in the oscillation period with increasing height.

\par Magnetic field configuration is a probable cause for such a distribution of oscillations in faculae. However, the magnetic field azimuthal angle in faculae seems somewhat random \citep{pillet}, and the connection between the oscillations at different height levels is difficult to trace \citep{kobanov11b}.

\par In this work we analyse characteristics of intensity, Doppler velocity and magnetic field oscillations in facular magnetic knots and compare them to those in facular patches with magnetic field of a moderate strength.

\section{Data and Methods}
\par For this research we used data provided by the Solar Dynamics Observatory (SDO). Instrument AIA provides full-disk images in 10 channels, of which we used three: the continuum 1700\,\AA, He\,\textsc{ii}\,304\,\AA, and Fe\,\textsc{ix}\,171\,\AA\ formed in the upper photosphere, transition region, and lower corona, respectively. Also, we used HMI Dopplergrams, magnetograms, and intensity images in the Fe\,\textsc{i}\,6173\,\AA\ line, which forms in the photosphere at a height of 200\,km \citep{parnell}. The pixel size corresponds to 0.6\arcsec\ for AIA and to 0.5\arcsec\ for HMI. The time cadence of the AIA data is 12 and 24\,s, and that of HMI data is 45\,s. 1.5 level data were used prepared with the \textit{aia\_prep} IDL routine. When needed, the co-alignment between the channels was corrected using sunspots at the solar disk as reference points. The full-vector magnetic field parameters, specifically the field inclination, were retrieved from the Milne-Eddington inversion data available at the JSOC \citep{Hoeksema}. The maps show inclination to the line of sight, which differs from the inclination to the solar radius by no more than 10\degree, since the faculae were located close to disk centre.

\par
\begin{figure}
\centerline{
\includegraphics[width=9cm]{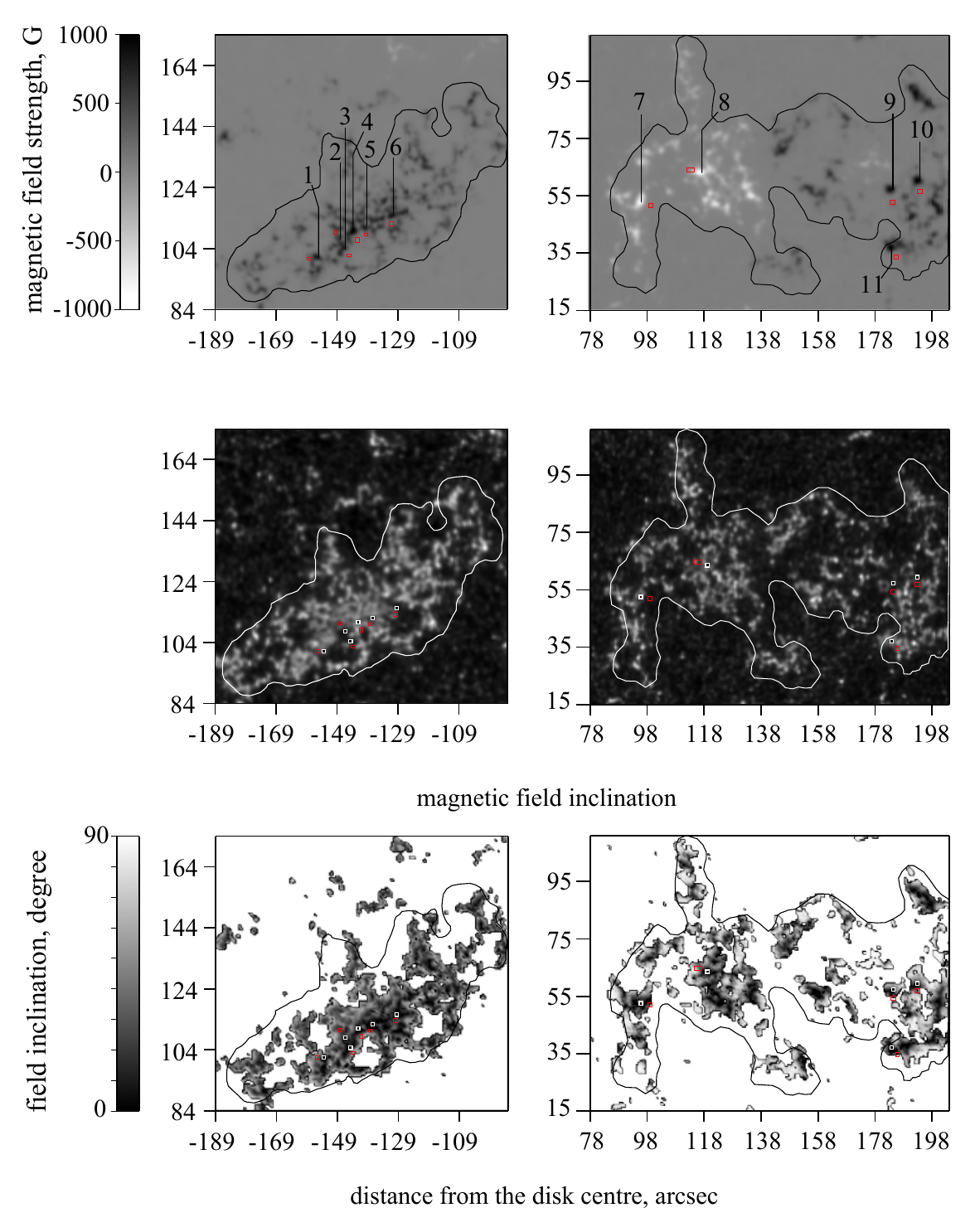}
}
\caption{\textit{Upper row:} distribution of the magnetic field in the researched faculae averaged over the series duration time. \textit{Middle row:} the boundaries of the faculae were defined using the 1700\,\AA\ SDO/AIA channel based on the images averaged over the series time. \textit{Lower row:} magnetic field inclination to the line of sight in the Fe\,\textsc{i}\,6173\,\AA\ line based on the first image of the series . Only the pixels whose magnetic field strength exceeds 70\,G are displayed in the panel, since for the lower strength pixels the error is far too big. The magnetic knots are labelled with numbers. The red rectangles indicate the patches within faculae that were taken as moderate field comparison regions.}
\label{fig:1}
\end{figure}

 As an object for the research we picked two faculae located close to disk centre observed on August\,14, 2010 and October\,1, 2011, the series lengths are 90 and 85\,min (Figure\,\ref{fig:1}).

\par We determined the facula boundaries as 0.7 peak brightness level of the 1700\,\AA\ image averaged over the series time blurred by 20 pixels (Figure\,\ref{fig:1}, middle panel). The slow trend was removed subtracting a smoothed signal from the original signal. For the analysis, we used the Fast Fourier Transform (FFT) IDL algorithm; to estimate statistical significance, we used the algorithm described in \citet{Torrence}.

\section{Results} 
      \label{S-general}
\subsection{Oscillations and Fan Structures Above Sunspots}



\par We determine facular magnetic knots as regions within facula boundaries whose magnetic field strength does not drop below 600\,G during the time series. As a rule, such regions are located in the central part of a facula, and their size is 1\arcsec\ to 2.5\arcsec. As comparison regions we used moderate-field regions within the faculae, whose magnetic field strength was 100--300\,G. Both the magnetic knots and comparison regions are shown in Figure\,\ref{fig:1}.

\par
\begin{figure}
\centerline{
\includegraphics[width=9cm]{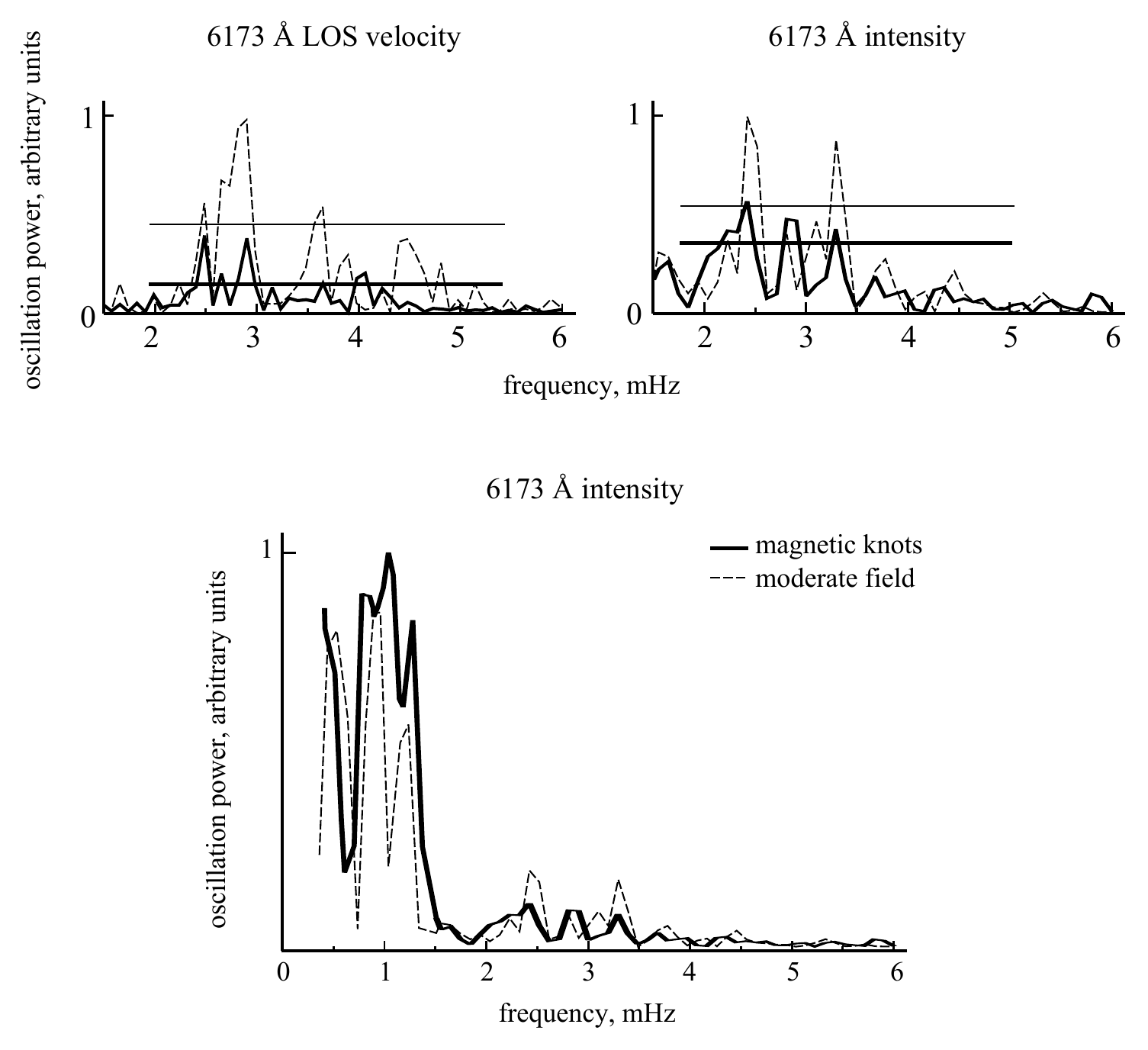}
}
\caption{HMI velocity and intensity spectra in magnetic knot 4 and those in a moderate-field patch. The horizontal lines indicate 95\% significance levels. The lower panel shows the same spectra with the frequency scale extended down to zero.}
\label{fig:2}
\end{figure}

 We restricted our main analysis to the 1.5--6\,mHz range. Oscillation power of the LOS velocity and magnetic field strength signals is noticeably lower than the significance level beyond this range. In addition, this excludes low-frequency oscillations, which dominate in intensity signals and impede higher-frequency components analysis. This can be seen in the lower panel of the Figure\,\ref{fig:2}, which shows an expanded spectrum of the intensity oscillations in the Fe\,\textsc{i}\,6173\,\AA\ line. In this range, the spectra of the oscillations at the lowest level of the solar atmosphere available---photospheric signals of the HMI Fe\,\textsc{i}\,6173\,\AA\ line---show no significant difference between magnetic knots and peripheral facular regions. This similarity in the spectral composition is characteristic of both intensity and Doppler-velocity oscillations. The oscillation power, however, is different: the oscillations in magnetic knots are on average 2--3 times weaker than that in the medium-field regions, again, in both the intensity and Doppler-velocity signals (Figure\,\ref{fig:2}). The moderate-field regions that we used to compare with the magnetic knots have the same sizes that magnetic knots have. To calculate the spectrum of a single magnetic knot or moderate-field region, here and further we averaged the signals over the knot area and then Fourier transformed the resulting signal.

\par
\begin{figure}
\centerline{
\includegraphics[width=10cm]{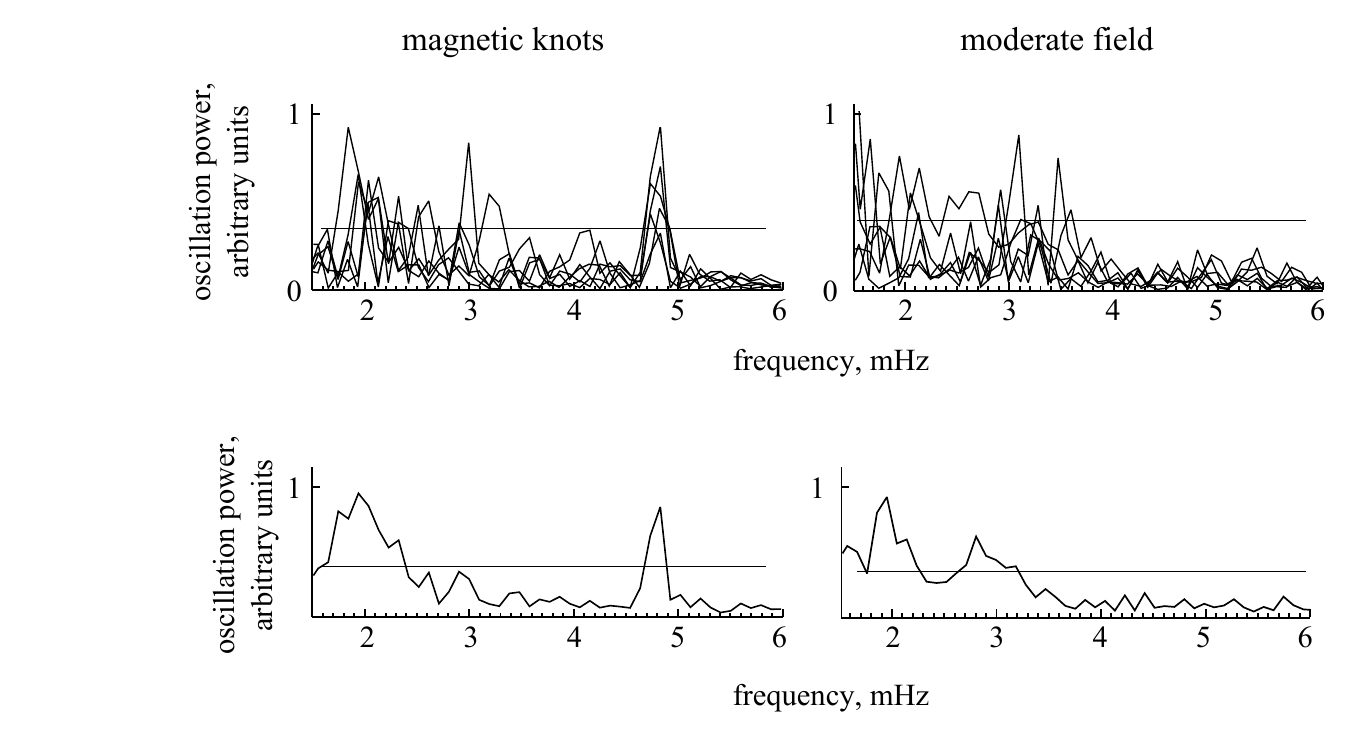}
}
\caption{Upper: examples of the magnetic field spectra in magnetic knots and in moderate-field regions; bottom: spectra of the magnetic field signals averaged over 11 knots and 11 moderate-field regions within the faculae. The horizontal lines indicate 95\% significance level.}
\label{fig:3}
\end{figure}

\begin{figure}
\centerline{
\includegraphics[width=11cm]{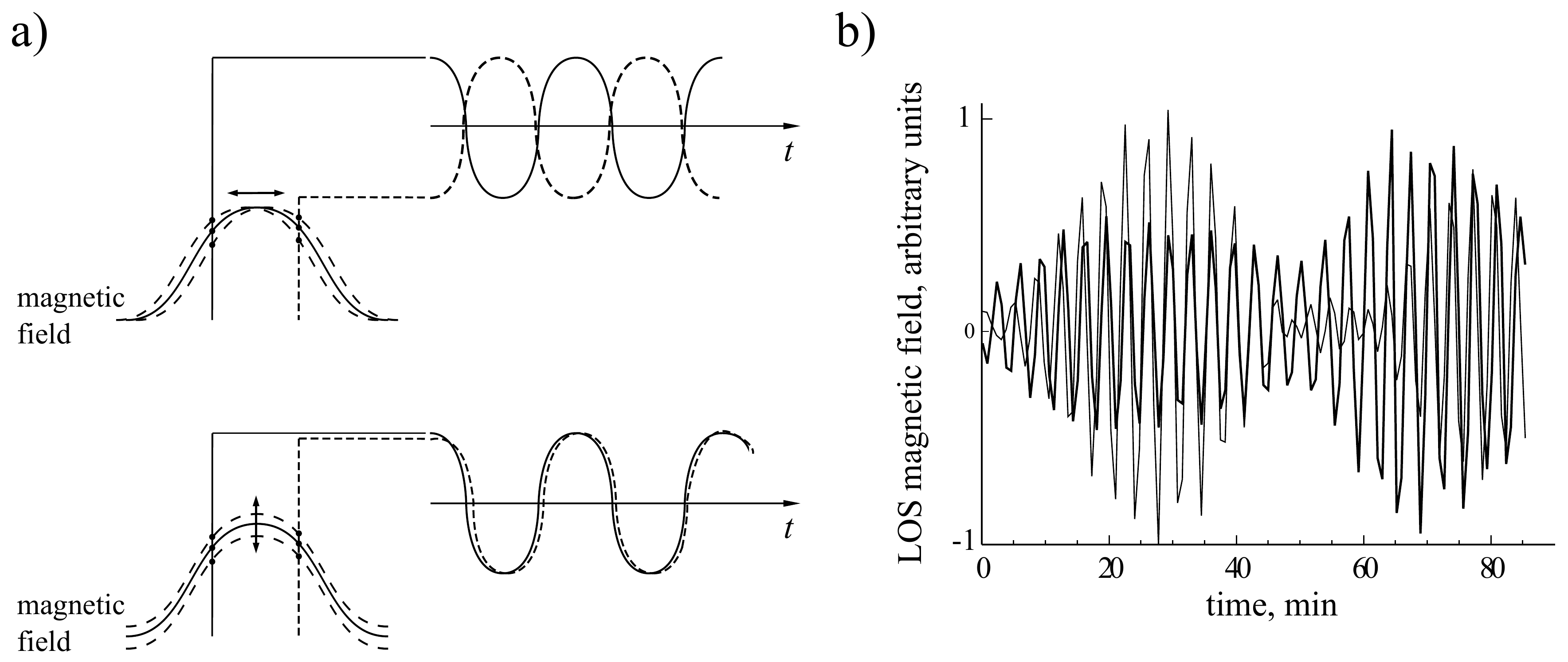}
}
\caption{\textit{a}: notion of the phase relation between the magnetic field signals around a magnetic knot in two scenarios  of the variation origin; \textit{b}: an example of magnetic field signals of the points located at opposite sides of magnetic knot\,4. The signals are filtered in the three-minute range.}
\label{fig:4}
\end{figure}

 Magnetic field strength spectra demonstrate a more noticeable difference: there are significant peaks in magnetic-knot spectra at a frequency of 4.8$\pm$0.1\,mHz. Figure\,\ref{fig:3} shows individual magnetic field oscillation spectra for the magnetic knots as well as averaged ones compared with those for the moderate-field regions. The question on these oscillations arose: are they real oscillations of the magnetic flux, or are they a product of magnetic hill horizontal movements, which may be caused by, e.g., horizontal waves? To test these hypotheses, we compared oscillations of the points placed at the opposite sides of magnetic hills. This analysis is based on the assumption that in the case of the horizontally moving hill the resulting signals would have opposite phases, even if the horizontal movements were of a sub-resolution scale. If the oscillations were caused by the real magnetic-field strength variations, the signals would be in-phase (Figure\,\ref{fig:4},\textit{a}). The overwhelming majority of the observed signal pairs show similar phases(Figure\,\ref{fig:4},\textit{b}), which led us to the conclusion that the oscillations are indeed real oscillations of the magnetic flux.

\par
\begin{figure}
\centerline{
\includegraphics[width=11cm]{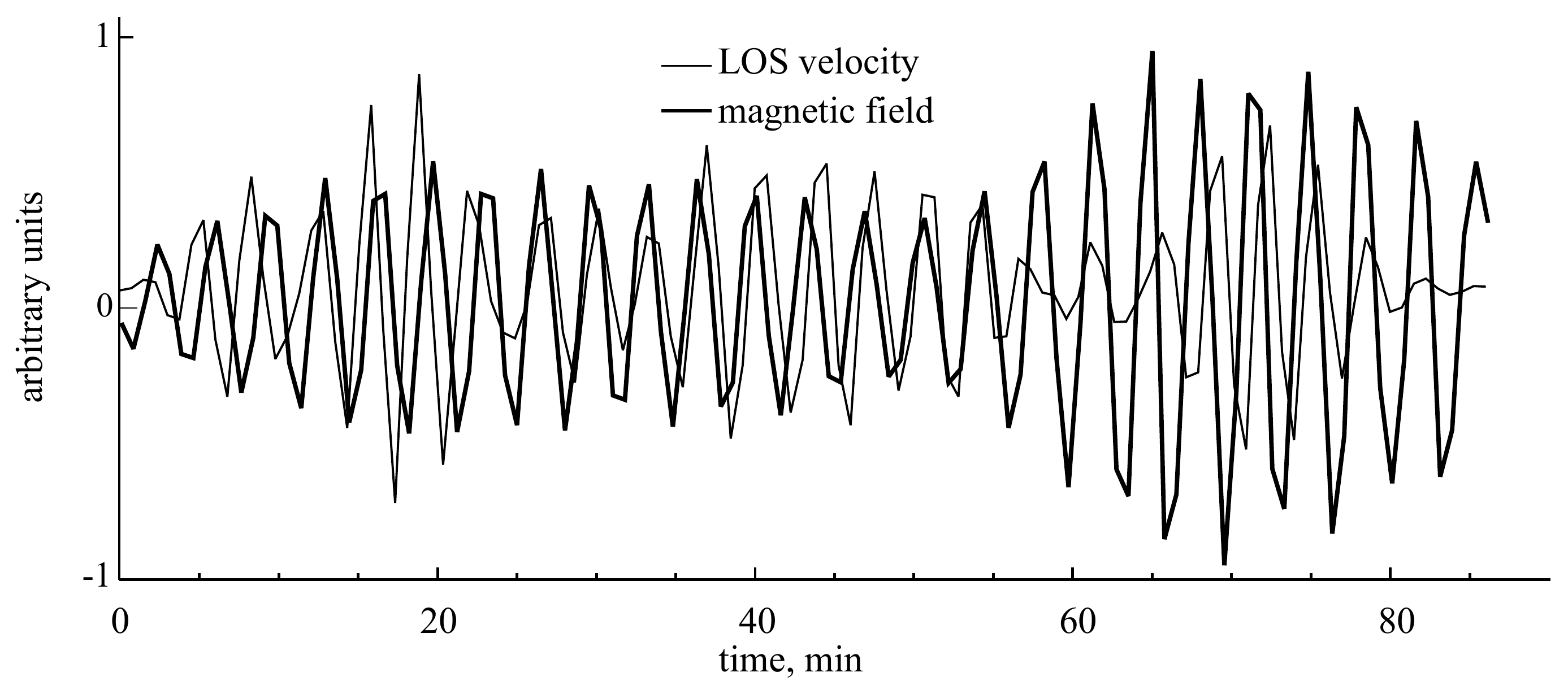}
}
\caption{Oscillations of the magnetic field strength and line-of-sight velocity in magnetic knot\,4 filtered in the three-minute range.}
\label{fig:9}
\end{figure}

 To examine the possible reasons for these oscillations, we compared them with the velocity oscillations. Figure\,\ref{fig:9} demonstrates that both the oscillation phase and the wave-train structure of the velocity oscillations show no correlation with those of magnetic field oscillations. Although Figure\,\ref{fig:9} is only for knot\,4, this result is typical for all the studied magnetic knots. Earlier, \citet{muglach} and \citet{kobanov07} as well registered real oscillations of the magnetic flux in facular regions.

\par Here, the term "real oscillations" implies that the observed oscillations are of the solar origin rather than a result of instrumental artefacts. In the solar origin signals we also include the possible cross-talk of the quasi-periodic plasma movements into the magnetic field signal. There are several scenarios for such a cross-talk: 1) small-scale spatial unresolved movements can cause quasi-periodic variations of the Stokes-V; 2) large-scale sub-photospheric and photospheric movements can make magnetic tube direction periodically divert near the line of sight, which, in turn, results in measured magnetic field signal variations, even given the fixed spatial location of the observed object; 3) density variations caused by the p-mode lead to the periodic changes in the line formation height, which ends up in the magnetic field signal variations, given the vertical gradient in the magnetic field. One should note that the cross-talk problem is relevant for all the instruments, both ground- and satellite-based. Analysing SOHO/MDI magnetograms, \citet{Norton} found magnetic field oscillations in a sunspot in three frequency ranges: 0.5--1.0, 3.0--3.5, 5.5--6.0\,mHz. Later \citet{Settele} showed that the observed oscillations result from the velocity signal cross-talk. \citet{Liu} claimed that such cross-talk is absent in the HMI magnetograms. However, as follows from the above, the situations does not seem so 'unclouded'. The investigation of the magnetic field variation sources in more detail constitutes a separate problem, which is out of scope of this work.

\par
\begin{figure}
\centerline{
\includegraphics[width=10cm]{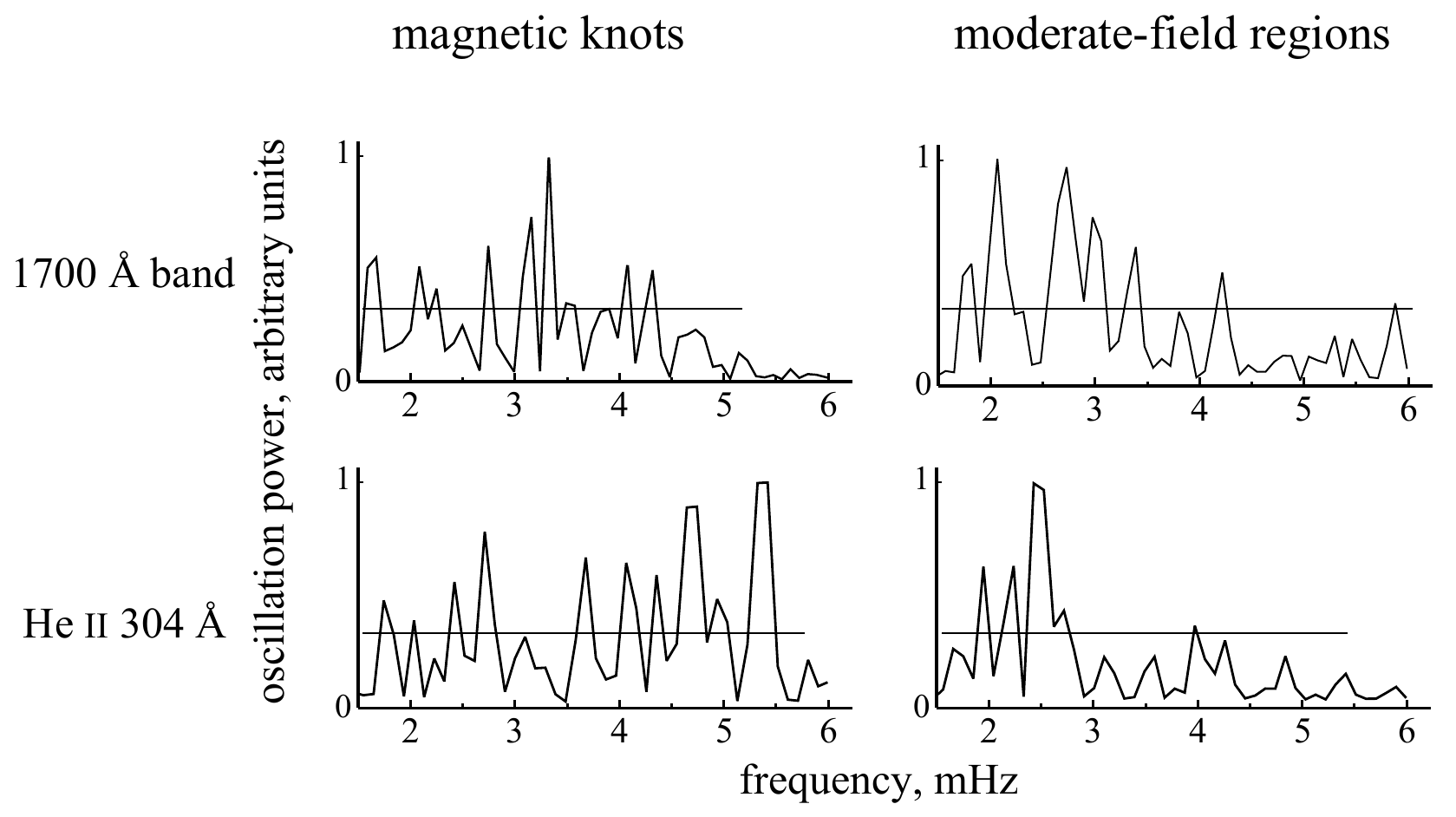}
}
\caption{The 1700\,\AA\ and 304\,\AA\ line intensity oscillation spectra in magnetic knot\,2 and in the adjacent area. The horizontal lines indicate 95\% significance level.}
\label{fig:5}
\end{figure}

 At the next height level that we studied---the 1700\,\AA\ spectral band forming in the upper photosphere---magnetic-knot spectra peaks show a slight shift to the high-frequencies range compared to the moderate-field spectra (Figure\,\ref{fig:5}, upper row). The He\,\textsc{ii}\,304\,\AA\ line intensity spectra representing the transition region of the solar atmosphere differ more noticeably: the significant peaks in the magnetic knot oscillation spectra are concentrated in the 3--6\,mHz range, and those of the peripheral-region oscillation spectra are distributed below 3\,mHz (Figure\,\ref{fig:5}, lower row).

\par
\begin{figure}
\centerline{
\includegraphics[width=9cm]{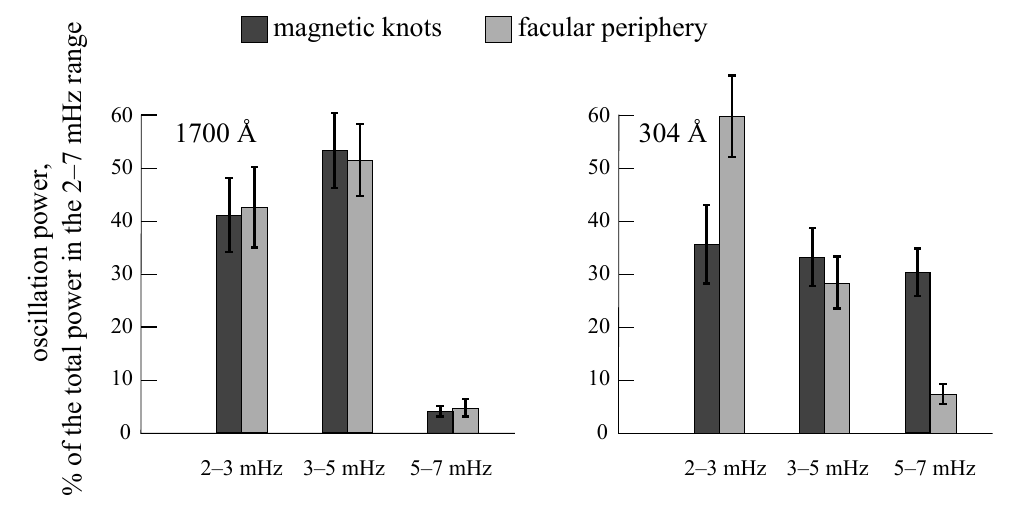}
}
\caption{Integral oscillation power in three frequency ranges in percent of the total 2--7\,mHz range oscillation power averaged over all the 11 magnetic knots compared with 11 peripheral regions in the two faculae with the error bars showing standard deviation over all the studied regions. In the diagrams only peaks counted whose power exceeds the 95\% significance level threshold.}
\label{fig:6}
\end{figure}

 The diagram in Figure\,\ref{fig:6} is based on the data on all the magnetic knots in the two faculae. It gives a summary of the oscillation power distribution in the upper photosphere and transition region. The power is given in percentage of the total 2--7\,mHz oscillation power. When constructing this diagram, only peaks exceeding the 2-$\sigma^2$ level were used. Again, note that only frequencies above 2\,mHz are used in the analysis, since lower frequencies would take a large fraction of the integral oscillation power were they put in the diagrams (Figure\,\ref{fig:2}, lower panel).

\par
\begin{figure}
\centerline{
\includegraphics[width=10cm]{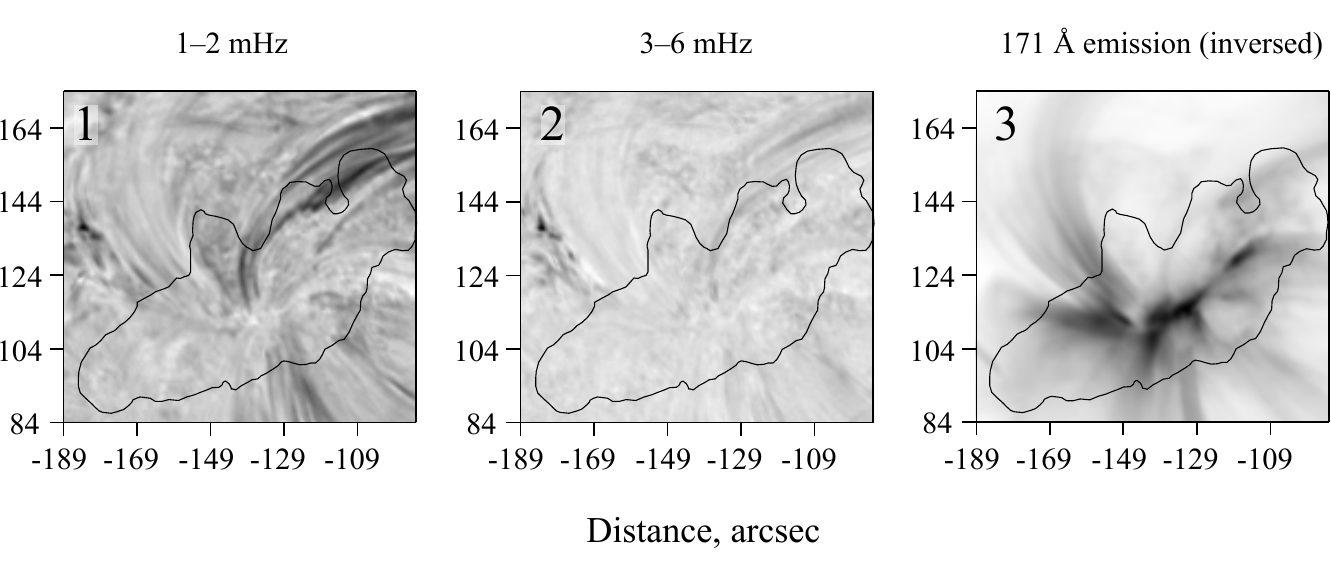}
}
\caption{\textit{Panels\,1 and 2}: spatial distributions of the 1--2\,mHz and 3--6\,mHz ranges oscillation power in the Fe\,\textsc{ix}\,171\,\AA\ line. Each element of the spatial distributions is total power of the Fourier transformed signals within the given frequency range. To minimize the influence of the brightness difference between the loops and the background, we subtracted the mean values from each pixel signal and normalized the resulting oscillations to the original signal brightness. Dark indicates high oscillation power. \textit{Panel\,3}: the facula in the time-averaged Fe\,\textsc{ix}\,171\,\AA\ line emission with the marked magnetic knot locations. The black and white colours are inversed so that the loops could be seen better.}
\label{fig:8}
\end{figure}

 Previously we noted that low frequency oscillations dominate above faculae in the lower corona observed in the Fe\,\textsc{ix}\,171\,\AA\ line \citep{Kobanov15, kobanov14}. The spatial distributions of the low-frequency oscillation power outline loop and fan structures seen in the coronal emission lines (Figure\,\ref{fig:8}). Most of the seen loops seem to have footpoints anchored approximately in the magnetic knots, though the co-alignment is the least precise for the Fe\,\textsc{ix}\,171\,\AA\ line.

\par The observed spectral characteristics in the photosphere and transition region can be explained by magnetic field configuration in faculae. At the height of 200\,km, where the Fe\,\textsc{i}\,6173\,\AA\ line forms \citep{parnell}, the magnetic flux is closer to vertical in both strong and medium field strength regions. Higher, the field lines begin to divert from the vertical more significantly, and the inclination is greater at the periphery of a facula than at its centre. This allows longer-period waves to propagate upward at the periphery, since the cut-off frequency depends on the angle between the magnetic field and the wave vector, which influences the transformation between the MHD modes and may form magneto-acoustic portals for low-frequency wave propagation \citep{Bel,Schunker,Jefferies,McIntosh}.

\par One could suggest that, as well as sunspots, each magnetic knot is surrounded by a magnetic field canopy. It would be natural to expect that in this case the spatial distribution of the dominant frequencies had the characteristics mentioned in the papers \citet{Moretti, Krijger, mugl}. In faculae, however, canopies of the chaotically located magnetic knots intersect forming a more complicated magnetic field topology. This impedes the unambiguous interpretation of the 2-d distributions of the dominant frequencies in faculae. The task looks even more complex when the analysis is carried out for several levels of the solar atmosphere. This is why in this work we restricted to the comparison analysis of the oscillation spectral characteristics in magnetic knots and the adjacent moderate field regions.

\par Note that our results do not agree with those of Kostik and Khomenko (2013), who also studied the dependence of the oscillation characteristics on magnetic field strength in faculae in the Ba\,\textsc{ii}\,4554\,\AA\ and Ca\,\textsc{ii}\,H lines. The authors of this study observed an increase of 15--20\,\% in the photospheric and chromospheric dominant oscillation period with the increase in the field strength. They, however, used regions with the vector field strength ranging from 500\,G to 1600\,G, while in our analysis the longitude field strength only reaches 1\,kG values, which could be somewhat an effect of the lower spatial resolution of the data that we used. In addition, the data series in our analysis are 2--3 times longer than the series that Kostik and Khomenko (2013) used. This provided a higher frequency resolution and allowed us to extend our analysis to the lower frequencies. We can not exclude the possibility that the found differences could be due to different properties of the spectral lines used in these studies.

\par The aforementioned suggests that the stated problem exists, and solving it requires further research in a wide spectral range with a greater sets of observational data.

\section{Conclusions} 
      \label{S-Conclusions}
\par i) Magnetic field strength oscillation power spectra in facular magnetic knots have prominent peaks at a frequency of about 4.8\,mHz, while spectra of the moderate strength patches lack such peaks. ii) The intensity and Doppler velocity spectral composition of the photospheric Fe\,\textsc{i}\,6173\,\AA\ line is constant over the area of faculae. In magnetic knots, however, the oscillation power is reduced 2--3-fold in both intensity and velocity signals. iii) At the upper-photospheric level of the 1700\,\AA\ line, the dominant frequency of the oscillations in knots is increased by 0.5--1\,mHz compared to the moderate-field patches. iv) In the transition region---the He\,\textsc{ii}\,304\,\AA\ line---the significant peaks of the magnetic knot oscillation spectra sit between 3 and 6\,mHz, and those of the peripheral region spectra sit below 3\,mHz.

\par We believe that such parameters of the oscillation distributions are caused by the magnetic field configuration in faculae. At the low atmospheric heights, the magnetic field lines are close to vertical over the whole area of a faculae, and the spectral composition at these heights does not change significantly in different parts of a facula. The field lines are inclined at the facular periphery in the upper photosphere, and more so in the transition region. This causes the decrease in the cut-off frequency and enables low-frequency waves to propagate upward.

\hyphenation{Pro-jects}
\small{\textbf{Acknowledgements}.
This study was supported by Project No.\,16.3.2 of ISTP SB RAS, by the Russian Foundation for Basic Research under grants 16-32-00268 mol\_a and No.\,15-32-20504 mol\_a\_ved. We acknowledge the NASA/SDO science team for providing
the data. We are grateful to the highly skilled anonymous reviewer, whose valuable remarks and suggestions helped us greatly improve the paper. 


\bibliographystyle{spr-mp-sola}

\tracingmacros=2
\bibliography{chelpanov-16}


\end{document}